\documentclass[prb,aps,epsfig,twocolumn,showpacs]{revtex4-1}
\usepackage{color}
\usepackage{epsfig,amssymb,amsmath,latexsym}
\usepackage{float}
\usepackage[colorlinks=true,linktoc=page,linkcolor=blue,citecolor=magenta]{hyperref}

\newcommand\bea{\begin{eqnarray}}
\newcommand\eea{\end{eqnarray}}
\newcommand\beq{\begin{equation}}  
\newcommand\eeq{\end{equation}}

\newcommand{\non}{\nonumber}  

\newcommand\ie{{\it{i.e.}}}
\newcommand\etal{{\it{et al.}}}
\newcommand\eg{{\it{e.g.}}}

\begin{document}
\textheight=23.8cm

\title{Quantum charge pumping through resonant crossed Andreev reflection in superconducting hybrid junction of Silicene}
\author{Ganesh C. Paul$^{1,2}$and Arijit Saha$^{1,2}$}
\affiliation{\mbox{$^1$} {Institute of Physics, Sachivalaya Marg, Bhubaneswar, Orissa, 751005, India} \\
\mbox{$^2$}{Homi Bhabha National Institute, Training School Complex, Anushakti Nagar, Mumbai 400085, India}
}

\date{\today}

\begin{abstract}
We theoretically investigate the phenomena of adiabatic quantum charge pumping through a normal-insulator-superconductor-insulator-normal (NISIN) 
setup of silicene within the scattering matrix formalism. Assuming thin barrier limit, we consider the strength of the two barriers 
($\chi_{1}$ and $\chi_{2}$) as the two pumping parameters in the adiabatic regime. 
Within this geometry, we obtain crossed Andreev reflection (CAR) with probability unity in the $\chi_{1}$-$\chi_{2}$ plane without 
concomitant transmission or elastic cotunneling (CT). Tunability of the band gap at the Dirac point by applying an external electric field 
perpendicular to the silicene sheet and variation of the chemical potential at the normal silicene region, open up the possibility of achieving 
perfect either CAR or transmission process through our setup. This resonant behavior is periodic with the barrier strengths.
We analyze the behavior of the pumped charge through the NISIN structure as a function of the pumping strength and angles of the 
incident electrons. We show that large ($Q\sim2e$) pumped charge can be obtained through our geometry when the pumping 
contour encloses either the CAR or transmission resonance in the pumping parameter space. We discuss possible experimental feasibility 
of our theoretical predictions.
\end{abstract} 
\maketitle

\section{Introduction}
In recent years, a close cousin to graphene~\cite{geimreview,castronetoreview}, silicene~\cite{liu2011low,houssa2015silicene,ezawa2015,ytanaka2016,
kaloni2016,siliceneexp1,siliceneexp2,siliceneexp3,lin2012structure} consisting of a monolayer honeycomb structure of silicon atoms, has attracted a 
lot of research interest in condensed matter community due to its unique Dirac like band structure which allows one to realize a rich varity of 
topological phases~\cite{CCLiu2,MEzawa1,ezawa2013spin,MEzawa5,ezawa2015antiferromagnetic,kaloni2014} and Majorana fermion~\cite{ezawa2015antiferromagnetic} 
in it under suitable circumstances. Moreover, this band structure is shown to be tunable by an external electric field applied perpendicular to the 
silicene sheet~\cite{drummond2012electrically,ezawa2012topological}. Dirac fermions, in turn, becomes massive at the two valleys ${\bf K}$ and 
${\bf K\rq{}}$ in this material. These properties have enable silicene to be a promising candidate for realizing spintronics~\cite{RevModPhys.76.323,wang2012half,
wang2015silicene,tsai2013gated,MEzawa6}, valleytronics~\cite{ezawa2013spin,pan2014valley,TYokoyama,saxena2015} devices as well as silicon based 
transistor~\cite{tao2015silicene} at room temperature.

Very recently, superconducting proximity effect in silicene has been investigated theoretically in Ref.~\onlinecite{Linder2014,paul2016thermal,
surajit2016conductance}. Although, up to now, no experiment has been put forwarded in the context of proximity effect in silicene.
In Ref.~\onlinecite{Linder2014}, a unique possibility of acquiring pure crossed Andreev reflection (CAR) without any contamination 
from normal transmission/co-tunneling (CT) has been reported in normal-superconductor-normal (NSN) junction of silicene where elastic cotunneling as well as Andreev reflection can be suppressed to zero by properly tuning the chemical potential and band gap at the two normal sides. However, in such NSN junction, maximum value of CAR probability does not reach 100\% because normal reflection does not vanish.
This naturally motivates us to study a NISIN junction of 
silicene and explore whether incorporating  an insulating barrier at each NS interface can give rise to resonant CAR in such setup.

On the other hand, adiabatic quantum pumping, 
is a transport phenomena in which low-frequency periodic modulations of at least two system parameters~\cite{thouless1983quantization,
buttiker1994current,brouwer1998scattering,brouwer2001rectification} with a phase difference lead to a zero bias finite dc current in meso and 
nanoscale systems. Such zero-bias current is obtained as a consequence of the time variation of the parameters of the quantum system, which 
explicitly breaks time-reversal symmetry~\cite{moskalets2002floquet,moskalets2004adiabatic,kundu2011quantum}. It is necessary to 
break time-reversal symmetry in order to  get net pumped charge, but it is not a sufficient condition. Indeed, in order to obtain a finite net 
pumped charge, parity or spatial symmetry  must also be broken. Finally, to reach the adiabatic limit, the required condition to satisfy is that
the period $T$ of the oscillatory driving signals has to be much larger than the dwell time $\tau_{dwell}\simeq L/v_F$ of the electrons 
inside the scattering region of length $L$, \ie, $T = 2\pi/ \omega\gg\tau_{dwell}$~\cite{brouwer1998scattering}. In this limit, the pumped 
charge in a unit cycle becomes independent of the pumping frequency. This is referred to as ``adiabatic quantum charge pumping''~\cite{brouwer1998scattering}.

During the past decades, quantum charge and spin pumping has been studied extensively in mesoscopic setups including quantum dots and 
quantum wires both at the theoretical~\cite{niu1990towards,niu1990towards,aleiner1998adiabatic,shutenko2000mesoscopic,moskalets2002floquet,
moskalets2004adiabatic,entin2002adiabatic,entin2002quantized,colin2004,colin2005,das2005effects,banerjee2007adiabatic,amitagarwal2007,amitagarwal2008,
tiwari2010quantum,RuiZhu,jurgen2008,jurgenreview2014} as well as experimental~\cite{switkes1999adiabatic,leek2005charge,watson2003experimental,
buitelaar2008adiabatic,giazotto2011josephson,blumenthal2007gigahertz} level with focus on both the adiabatic and nonadiabatic regime. 
In recent times, quantum pumping has been explored in Dirac systems like graphene~\cite{RuiZhu,prada2009quantum,tiwari2010quantum,kundu2011quantum,
alos2011adiabatic} and topological insulator~\cite{citro2011electrically,rakeshTIpumping2012}. However, the possible quantization of 
pumped charge~\cite{avr01} during a cycle through non-interacting open quantum systems has been investigated so far based on the resonant 
transmission process~\cite{lev,entin2002quantized,kundu2011quantum,saha2014quantum}. In more recent times, quantized behavior of pumped charge
has been predicted in superconducting wires with Majorana fermions~\cite{gib13}, fractional fermions~\cite{saha2014quantum} and topological 
insulators in enlarged parameter spaces~\cite{PedroLopes}. Although, till date, quantum pumping phenomena through resonant CAR process has not 
been investigated to the best of our knowledge.

Motivated by the above mentioned facts, in this article, we study adiabatic quantum charge pumping either through resonant CAR process or 
resonant transmission process, under suitable circumstances, in silicene NISIN junction. We model our pump setup within the scattering matrix
formalism~\cite{buttiker1994current,brouwer1998scattering} and consider the strength of the two barriers (in the thin barrier limit)
as our pumping parameters. We show that CAR probability can be unity in the pumping parameter space. Moreover, resonant CAR is periodic
in the pumping parameter space due to the relativistic nature of the Dirac fermions. Similar periodicity is present, in case of 
resonant tunneling process as well, under suitable condition. Adiabatic quantum pumping through these processes, with the modulation 
of two barrier strengths, can lead to large pumped charge from one reservoir to the other. We investigate the 
nature of pumped charge through NISIN structure as a function of the pumping strength and angle of incidence of incoming electrons choosing 
different types of pumping contours (circular, elliptic, lemniscate~\cite{saha2014quantum} etc.).

The remainder of the paper is organized as follows. In Sec.~\ref{sec:II}, we describe our pump setup based on the silicene NISIN junction
and the formula for computing pumped charge within the scattering matrix framework. Sec.~\ref{sec:III} is devoted to the numerical results 
obtained for the pumped charge as a function of various parameters of the systems. Finally, we summarize our results and conclude 
in Sec.~\ref{sec:IV}.

\section{Model and Method} {\label{sec:II}}

\begin{figure}[!thpb]
\centering
\includegraphics[width=1.0\linewidth]{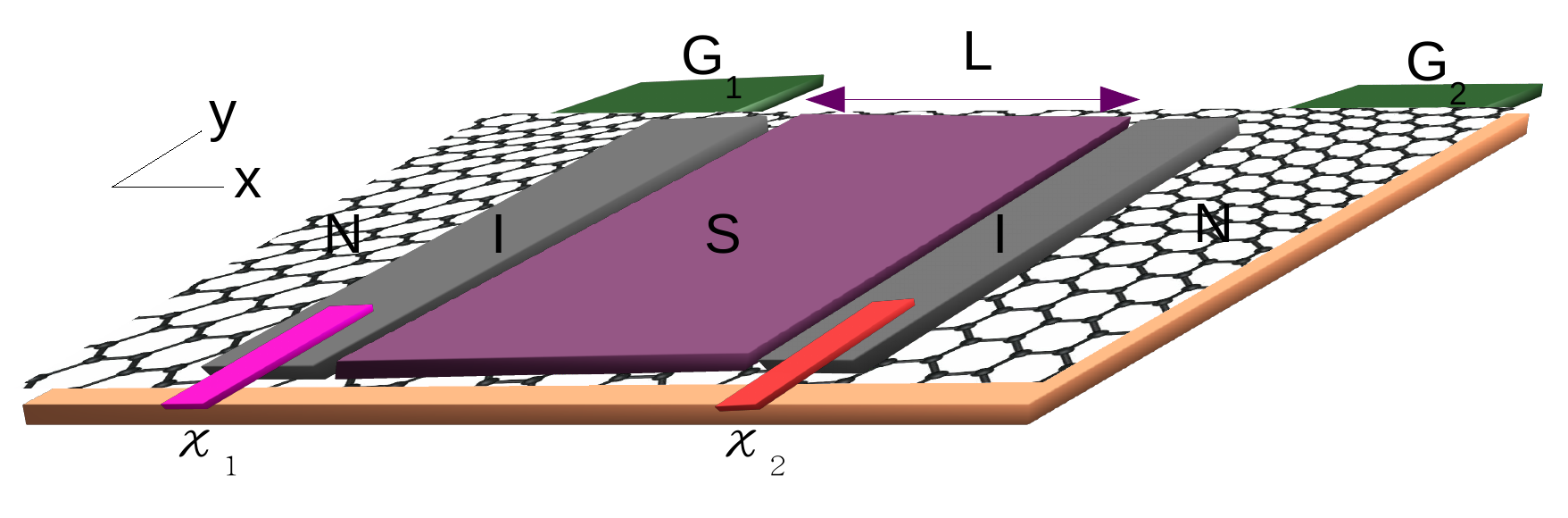}
\caption{(Color online) A schematic sketch of our silicene NISIN set-up. Silicene sheet with hexagonal lattice structure is deposited on 
a substrate (orange, light grey). Here $N$ indicates the normal region, $I$ denotes the thin insulating barrier region (grey, light grey).  
A bulk superconducting material of length $L$, denoted by $S$ (pink, light grey), is placed in close proximity to the silicene sheet to induce 
superconducting correlation in it. Two gates $G_1$ and $G_2$ (dark green, dark grey) are connected to the two normal regions (N) 
of the silicene sheet to tune the chemical potential (doping) there.
Two extra gates (blue and red, light grey) indicated by $\chi_{1}$ and $\chi_{2}$ are symbolically denoted 
to modulate the barrier strengths.}
\label{model}
\end{figure}

In this section we describe our pump setup in which we consider a normal-insulator-superconductor-insulator-normal (NISIN) structure of silicene in 
$x-y$ plane as depicted in Fig.~\ref{model}. Here, the superconducting region being located between $0 < x < L$, while the insulating barriers 
are situated on its left, $-d < x < 0$, and on its right, $L < x < L + d$. The normal region of silicene occupies at the extreme left \ie, $x < - d$ 
and extreme right ends, $x > L + d$. Here, superconductivity is assumed to be induced in the silicene sheet via the proximity effect, where a bulk 
$s$-wave superconductor is placed in close proximity to the sheet in the region $0 < x < L$. The two insulating regions in silicene have gate
tunable barriers of strength $\chi_{1}$ and $\chi_{2}$ in the thin barrier limit~\cite{paul2016thermal,surajit2016conductance}. 
Two additional gate voltages $G_{1}$ and $G_{2}$ can tune the chemical potential in the left and right normal silicene regions respectively.

The silicene NISIN junction can be described by the Dirac Bogoliubov-de Gennes (DBdG) equation of the form~\cite{Linder2014,paul2016thermal}

\begin{eqnarray}
\begin{bmatrix}
\hat{H}_{\tilde{\eta}} & \Delta \hat{1}\\
\Delta^{\dagger}\hat{1}  & -\hat{H}_{\tilde{\eta}}
\end{bmatrix} {\Psi}=E{\Psi}\ .
\label{bdgHamilt}
\end{eqnarray}

where $E$ is the excitation energy, $\Delta$ is the proximity induced superconducting pairing gap. The Hamiltonian $H_{\tilde{\eta}}$ describes the 
low energy physics close to each $\bf{K}$ and $\bf{K^{\prime}}$ Dirac points and reads as~\cite{ezawa2012topological}

\begin{align}
H_{\tilde{\eta}}=\hbar v_f(\tilde{\eta} k_x \hat{\tau}{_x}-k_y \hat{\tau}{_y})+(elE_z-\tilde{\eta}\sigma\lambda{_S}{_O})\hat{\tau}{_z}-\mu \hat{1}\ .
\label{hamilt}
\end{align}

where $v_f$ is the Fermi velocity of the electrons, $\mu$ is the chemical potential, $\lambda_{SO}$ is the spin-orbit term and $E_{z}$ is the 
external electric field applied perpendicular to the silicene sheet. Here $\tilde{\eta}=\pm 1$ denotes the $\bf{K}$ and $\bf{K\rq{}}$ valley.
In Eq.~(\ref{hamilt}), $\sigma$ is the spin index and $\hat{\tau}$ correspond to the Pauli matrices acting on the sub-lattices A and B where 
$\hat{1}$ is the $2\times2$ identity operator. 

The potential energy term $elE_{z}$ in the low energy Hamiltonian $H_{\tilde{\eta}}$ originates due to the buckled structure of silicene in which the 
A and B sublattices are non-coplanar (separated by a distance of length $l$) and therefore acquire a potential difference when an external 
electric field $E_{z}$ is applied perpendicular to the plane. It turns out that at a critical electric field $E^{c}_{z}=\lambda_{SO}/el$,
the band gap at each of the valleys become zero with the gapless modes of one of the valley being up-spin polarized and the other being 
down-spin polarised~\cite{drummond2012electrically,ezawa2012topological}. Away from the critical field, the bands (corresponding to $H_{\tilde{\eta}}$)  
at each of the valleys $\bf{K}$ and $\bf{K^{\prime}}$ split into two conduction and valence bands with the band gap being 
$|elE_{z} - \tilde{\eta}\sigma\lambda_{SO}|$. Note that, in silicene, the pairing occurs between $\tilde{\eta}=1$, $\sigma=1$ and $\tilde{\eta}=-1$, 
$\sigma=-1$ as well as $\tilde{\eta}=1$, $\sigma=-1$ and $\tilde{\eta}=-1$, $\sigma=1$ for a $s$-wave superconductor.

Here we set up the equations to analyze the quantum pumping phenomena through our NISIN structure. Solving Eq.(\ref{bdgHamilt}) we find the wave 
functions in three different regions. The wave functions for the electrons (e) and holes (h) moving in $\pm x$ direction in left or right normal 
silicene region $N$ reads

\begin{align}
\psi_{Nm}^{e{\pm}}= \frac{1}{A}
\begin{bmatrix}
 \frac{\pm{\tilde{\eta}}k^e_{1m}e^{\pm i\tilde{\eta}\alpha_{em}}}{\tau^e_{1m}}\\
 1\\
 0\\
 0
\end{bmatrix}
\exp[i({\pm}k^e_{1_{x}m}x+k^e_{1_{y}}y)] \ , \nonumber \\
\psi_{Nm}^{h{\pm}}= \frac{1}{B}
\begin{bmatrix}
 0\\
 0\\
 \frac{\mp{\tilde{\eta}}k^h_{1m}e^{\pm i\tilde{\eta}\alpha_{hm}}}{\tau^h_{1m}}\\
 1
\end{bmatrix}
\exp[i({\pm}k^h_{1_{x}m}x+k^h_{1_{y}}y)]\ .
\label{wfn}
\end{align}

where the index $m=\rm L/R$ stands for the left or right normal silicene region and we use this symbol for the rest of the paper. 
In Eq.(\ref{wfn}) the normalization factors are given by $A={\sqrt{\frac{2(E+{\mu_{m}})}{\tau^e_{1m}}}}$,\,\,$B={\sqrt{\frac{2(E-{\mu_{m}})}
{\tau^h_{1m}}}}$ and

\bea
k^{e(h)}_{1m}&=&\sqrt{\Big(k^{e(h)}_{1_{x}m}\Big)^2 + \Big(k^{e(h)}_{1_{y}}\Big)^2}\ ,
\eea

\begin{align}
k^{e(h)}_{1_{x}m}=\sqrt{(E{\pm}{\mu_{m}})^2 - (elE_{zm} - {\tilde{\eta}}{\sigma}\lambda_{SO})^2 - \Big(k^{e(h)}_{1_{y}}\Big)^2}\ .
\end{align}

\bea
{\tau^{e(h)}_{1m}}&=&E{\pm}{\mu_{m}}{\mp}(elE_{zm}-{\tilde{\eta}}{\sigma}\lambda_{SO})\ .
\eea

Here $\mu_{m}$ indicates the chemical potential in the left ($\mu_{L}$) or right ($\mu_{R}$) normal silicene region. $E$ is the energy of the 
incident particle. 

Due to the translational invariance in the $y$-direction, corresponding momentum $k^{e(h)}_{1_{y}}$ is conserved. Hence, the angle of incidence 
${\alpha_{em}}$ and the Andreev reflection (AR) angle ${\alpha_{hm}}$ are related via the relation 
\bea
k^h_{1m}\sin({\alpha_{hm}})=k^e_{1m}\sin({\alpha_{em}})\ . 
\eea

In the insulating region $I$, the corresponding wave functions can be inferred from normal region wave functions (Eq.(\ref{wfn})) by replacing 
$\mu_{m}\rightarrow \mu_{m}-V_0(V_0\rq{})$ where $V_0$ and $V_0\rq{}$ are the applied gate voltages at the left and right insulating regions 
respectively. We define dimensionless barrier strengths~\cite{paul2016thermal,surajit2016conductance} $\chi_1\,=\,V_0 d/\hbar v_F$ and 
$\chi_2\,=\,V_0\rq{} d/\hbar v_F$ which we use as pumping parameters for our analysis. Here $d$ is the width of the insulating barriers 
assumed to be the same for both of them.


\begin{figure}[!thpb]
\centering
\includegraphics[width=0.99\linewidth]{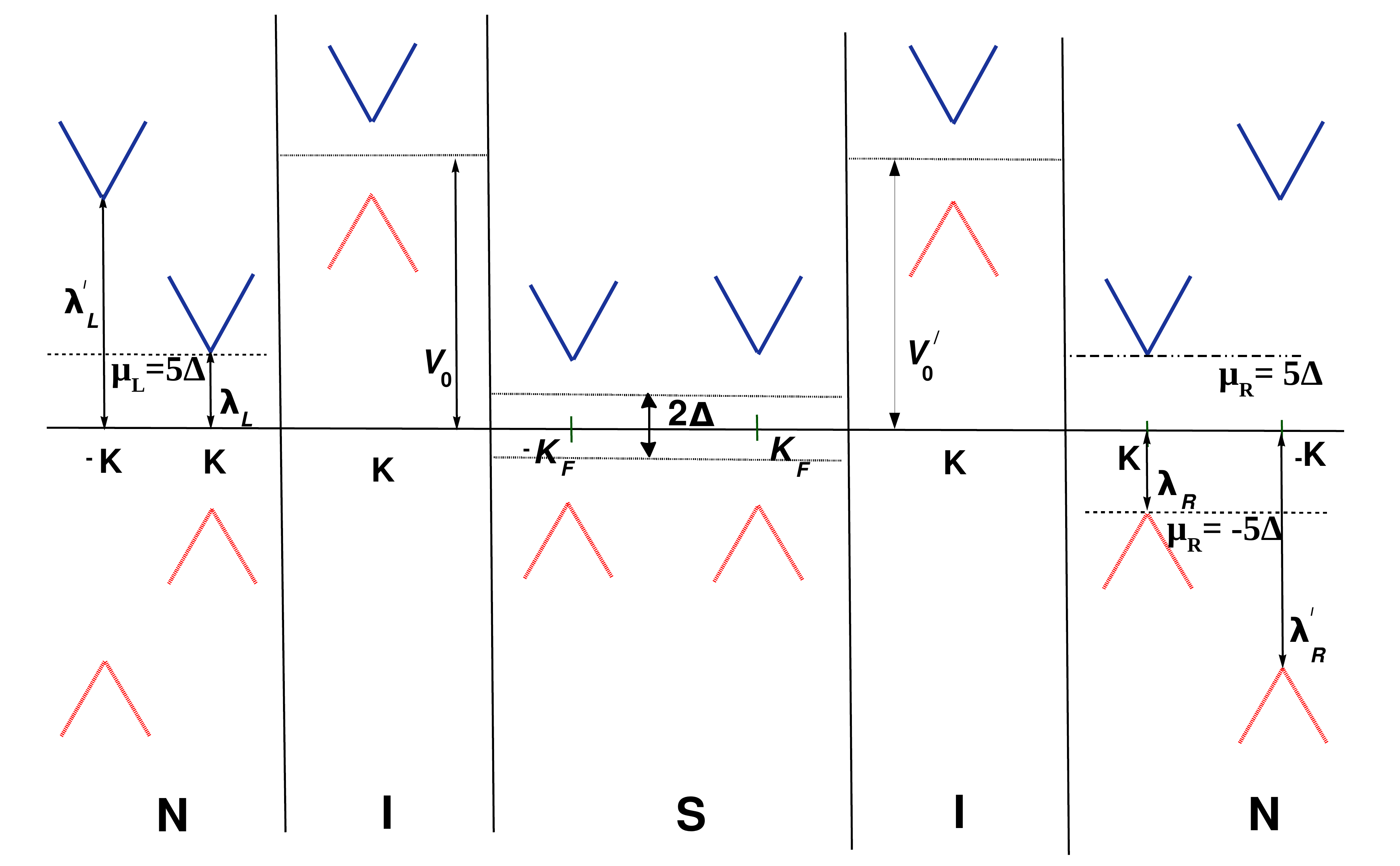}
\caption{(Color online) A schematic sketch of the band structure of our silicene NISIN setup is depicted. For the normal regions of 
silicene ($N$) as well as superconducting ($S$) silicene region, both ${\bf K}$ and ${\bf K\rq{}}$ valleys are presented. In contrast, 
only ${\bf K}$ valley is shown for both the insulating regions ($I$) for simplicity. 
Blue solid line indicates the conduction band while the valence bands are represented by red solid lines. 
At the right normal silicene side, the chemical potential is set at the top of the valance band ($\mu_R=-5\Delta$, dotted line) 
to obtain resonant CAR process. On the other hand, for resonant transmission to take place, chemical potential in the right normal side is set 
at the bottom of the conduction band ($\mu_R=5\Delta$, dotdashed line).
}
\label{band}
\end{figure}


In the superconducting region $S$, the wave functions of DBdG quasiparticles are given by,
\begin{eqnarray}
\psi_S^{e\pm} &=&
\frac{1}{\sqrt{2}}
\begin{bmatrix}
u_{1}\\
{\pm\tilde{\eta}}u_{1}e^i{^{\tilde{\eta}}}{^{\theta_e}}\\
u_{2}\\
{\pm\tilde{\eta}}u_{2}e^i{^{\tilde{\eta}}}{^{\theta_e}}
\end{bmatrix}
\exp[\pm(i{\mu}{_S}-{\kappa})x+iq^e_{y} y]\ , \notag\\
\psi_S^{h\mp}&=&\frac{1}{\sqrt{2}}
\begin{bmatrix}
u_{2}\\
{\mp\tilde{\eta}}u_{2}e^-{^i}{^{\tilde{\eta}}}{^{\theta_h}}\\
u_{1}\\
{\mp\tilde{\eta}}u_{1}e^-{^i}{^{\tilde{\eta}}}{^{\theta_e}}
\end{bmatrix}
\exp[\pm(-i{\mu}{_S}-{\kappa})x+iq^h_{y} y]\ .
\label{wfsc}
\end{eqnarray}

Here the coherence factors are given by, 
\begin{eqnarray}
u_{{1(2)}}={\Big[{\frac{1}{2}}\,{\pm}\,{\frac{\sqrt{E{^2}-{\Delta}{^2}}}{2E}}\Big]^{\frac{1}{2}}} {\rm and}~~~{\kappa}={\sqrt{{\Delta}{^2}-E^2}}.
\end{eqnarray}
As before, the translational invariance along the $y$ direction relates the transmission angles for the electron-like and hole-like quasi-particles 
via the following relation given by,
\bea
q{^{\beta}}\sin{\theta}_{\beta}=k^e_{1m} \sin{\alpha}_{em}\ .
\label{sc1}
\eea
for ${\beta}=e,h.$ The quasi-particle momentum can be written as
\bea
q^{e(h)}={\mu}{_S}\,{\pm}\,{\sqrt{E{^2}-{\Delta}{^2}}}\ .
\label{sc2}
\eea
where $\mu_S=\mu_m + U_0$, and $U_0$ is the gate potential applied to the superconducting region in order to tune the Fermi wave-length 
mismatch~\cite{beenakker1} between the normal and superconducting regions. The requirement for the mean-field treatment of superconductivity 
is justified in our model as we have taken $\mu_{S}\gg\Delta$~\cite{beenakker1,beenakkerreview} throughout our calculation.

We consider electrons with energy $E$ incident from the left normal region of the silicene sheet in the subgapped regime ($E<\Delta$). 
Considering normal reflection, Andreev reflection, cotunneling (normal transmission) and crossed Andreev reflection from the interface, 
we can write the wave functions in five different regions of the junction as

\begin{eqnarray}
\Psi_N^L&=&\psi_{NL}^{e+}+r_e\psi_{NL}^{e-}+r{_A}\psi_{NL}^{h-}\ ,\non\\
\Psi_I^L&=&p_1\psi_{IL}^{e+}+q_1\psi_{IL}^{e-}+m_1\psi_{IL}^{h+}+n_1\psi_{IL}^{h-}\ , \non\\
\Psi_S&=&t_{1}\psi_{S}^{e+}+t_{2}\psi_{S}^{e-}+t_{3}\psi_{S}^{h+}+t_{4}\psi_{S}^{h-}\ ,\non\\
\Psi_I^R&=&p_2\psi_{IR}^{e+}+q_2\psi_{IR}^{e-}+m_2\psi_{IR}^{h+}+n_2\psi_{IR}^{h-}\ , \non\\
\Psi_N^R&=&t_e\psi_{NR}^{e+}+t_A\psi_{NR}^{h+}\ .
\label{bc}
\end{eqnarray}

where $r_e$, $r_{A}$, $t_e$, $t_A$ correspond to the amplitudes of normal reflection, AR, transmission and CAR in the $N$ silicene regions, 
respectively. The transmission amplitudes $t_{1}$, $t_{2}$, $t_{3}$ and $t_{4}$ denote the electron like and hole like quasi-particles
in the $S$ region. Using the boundary conditions at the four interfaces, we can write

\begin{eqnarray}
\Psi_N^L|_{x=-d}&=&\Psi_I^L|_{x=-d},\,\,\,\,\,\, \Psi_I^L|_{x=0}=\Psi_S|_{x=0}\ , \non\\
\Psi_S|_{x=L}&=&\Psi_I^R|_{x=L},\,\,\,\,\,\,\, \Psi_I^R|_{x=L+d}=\Psi_N^R|_{x=L+d}\ .
\label{bc1}
\end{eqnarray}

which yields a set of sixteen linearly independent equations. Solving these equations numerically, we obtain $r_e$, $r_{A}$, $t_e$, $t_A$ 
which are required for the computation of pumped charge through our setup.

In order to carry out our analysis for the pumped charge in silicene NISIN structure, we choose the two dimensionless insulating barrier 
strengths $\chi_1$ and $\chi_2$ as our pumping parameters. They evolve in time either as (off-set circular contours)

\begin{eqnarray}
\chi_1=\chi_0 + P \cos(\omega t-\eta)\ ,\non\\
\chi_2=\chi_0 + P \cos(\omega t+\eta)\ ,
\label{cir}
\end{eqnarray}

or as (``lemniscate'' contours),

\begin{align}
\chi_{1}=\chi_{1_{0}}+P_{L}\Big(\cos{\theta}\cos{\omega t}-\frac{1}{2}\sin{\theta}\sin{2\omega t}\Big)/(1+\sin^2{\omega t})\ , \non\\
\chi_{2}=\chi_{2_{0}}+P_{L}\Big(\cos{\theta}\cos{\omega t}+\frac{1}{2}\sin{\theta}\sin{2\omega t}\Big)/(1+\sin^2{\omega t})\ ,
\label{lem}
\end{align}
respectively. In the circular contours $\chi_0$ and in the lemniscate contours $\chi_{1_{0}}$, $\chi_{2_{0}}$ correspond to the mean value of
the amplitude respectively, around which the two pumping parameters are modulated with time. $P$ and $P_{L}$ are called the pumping strengths 
for the two types of contours respectively. Furthermore, $2\eta$ and $\theta$ represent the phase offset between the two pumping signals for 
the circular and lemniscate contours, respectively. Here $\omega$ is the frequency of oscillation of the pumping parameters. 

We, in our analysis, only consider the adiabatic limit of quantum pumping where time period of the pumping parameters $T=2\pi/\omega$ is much longer 
than the dwell time $\tau_{dwell}\simeq L/v_F$ of the Dirac fermions inside the proximity induced superconducting region.

To calculate the pumped charge, we employ Brouwer's formula~\cite{brouwer1998scattering} which relies on the knowledge of the parametric 
derivatives of the $S$-matrix elements. Following Ref.~\onlinecite{kundusahaprb}, $S$-matrix for the NISIN structure of silicene for an incident 
electron with energy $E$, can be written as

\begin{eqnarray}
S=
\begin{bmatrix}
|r_e|e^{i{\gamma}_e} & |r_A| e^{i{\gamma}_h}& |t_e| e^{i\delta_e} & |t_A| e^{i\delta_h}\\
|r_A| e^{i{\gamma}_h} & |r_e| e^{i{\gamma}_e} & |t_A| e^{i\delta_h} & |t_e| e^{i\delta_e}\\
|t_e| e^{i\delta_e} & |t_A| e^{i\delta_h} & |r_e| e^{i{\gamma}_e} & |r_A| e^{i{\gamma}_h}\\
|t_A| e^{i{\delta}_h} & |t_e| e^{i{\delta}_e} & |r_A| e^{i{\gamma}_h} & |r_e| e^{i{\gamma}_e}
\end{bmatrix} \ ,
\label{Smatrix}
\end{eqnarray}

We write here the complex $S$-matrix elements $S_{ij}$ in polar form, with modulus and phase explicitly shown, since the
phase is going to play a major role in the determination of the pumped charge.
For a single channel $S$-matrix, the formula for the pumped charge becomes~\cite{kundusahaprb}

\begin{eqnarray}
Q=\frac{e}{2\pi}\int_0^{T} dt[&-&|r_A|^2 ({\dot{\gamma}}_h \cos{\alpha_{hL}}+\dot{\gamma}_e \cos{\alpha_{eL}})\non\\ 
&-&|t_A|^2(\dot{\delta}_h \cos{\alpha_{hR}}+\dot{\gamma}_e \cos{\alpha_{eL}})\non\\ 
&+&|t_e|^2(\dot{\delta}_e \cos{\alpha_{eR}}-\dot{\gamma}_e \cos{\alpha_{eL}})\non\\ 
&+&\dot{\gamma}_e \cos{\alpha_{eL}}]\ ,
\label{pumpch}
\end{eqnarray}

Here, we have redefined the complex scattering amplitudes $r_A$ and $t_A$ to satisfy the conservation of probability current
~\cite{Linder2014}. On the other hand, the other two scattering amplitudes $r_{e}$ and $t_{e}$ remain unchanged. Hence, the redefined scattering probabilities $|r_A|^2$ and $|t_A|^2$ become
\begin{eqnarray}
|r_A|^2 &\equiv&\frac{k^{h}_{1_{x}}}{k^{e}_{1_{x}}} \Bigg[\frac{2(E+\mu_L)(E-\mu_L-\lambda_L)}{|\eta k^{h}_{1_{x}}-i k^{e}_{1_{y}}|^2+
(E-\mu_L-\lambda_L)^2}\Bigg] |r_A|^2\ , \nonumber\\
|t_A|^2 &\equiv&\frac{k^{h}_{1_{x}}}{k^{e}_{1_{x}}} \Bigg[\frac{(E+\mu_L)}{(E-\mu_R)}\Bigg] |t_A|^2 \ .
\end{eqnarray}

Furthermore, ${\gamma}_e$, ${\gamma}_h$, $\delta_e$, $\delta_h$ are the phases of redefined $r_e$, $r_A$, $t_e$ and $t_A$ respectively. 
Here, $\alpha_{eL}$, $\alpha_{eR}$ correspond to the incident and transmitted angles of electrons while $\alpha_{hL}$, $\alpha_{hR}$ 
represent the reflected and transmitted angles of holes respectively. Note that, if $\alpha_{eL}=0$,
then the last term of Eq.(\ref{pumpch}) consisting of the time derivative of reflection phase is called 
``\textit{topological part}''~\cite{das2005effects} while the rest is termed as ``\textit{dissipative part}''~\cite{das2005effects}. 
The last term is called ``topological" becuase for $\alpha_{eL}=0$, it has to return to itself after the full period.
Hence, the only possible change in ${\gamma}_e$ in a period can be integer multiples of $2\pi$ \ie~${\gamma}_{e}(T)
\rightarrow {\gamma}_{e}(0) + 2\pi n$. On the other hand, the rest of the terms in Eq.(\ref{pumpch}) are together called ``dissipative" 
since their cumulative contribution prevents the perfect quantization of pumped charge. 

\section{Numerical Results}  {\label{sec:III}}
In this section we present and discuss our numerical results for the pumped charge based on Eq.(\ref{pumpch}). The quantum mechanical 
scattering amplitudes are all functions of the incident electron energy $E$, length of the superconducting silicene region $L$, 
the strengths $\chi_{1}$, $\chi_{2}$ of the two thin insulating barriers, chemical potential $\mu_{m}$ ($m=\rm L/R$) of the 
left and right normal silicene region, external electric field $E_{zm}$ ($m=\rm L/R$) and spin orbit coupling $\lambda_{SO}$. 
We denote the band gaps at the left and right normal silicene side as 2$\lambda_L$ and 2$\lambda_R$ 
respectively (see Fig.~\ref{band}) where $\lambda_m=(elE_{zm} - {\tilde{\eta}}{\sigma}{\lambda}{_S}{_O})$. In addition, 
we have set $\hbar=1$ throughout our analysis. 

For clarity, we divide this section into two subsections. In the first one, we discuss quantum pumping via resonant CAR process 
with unit probability in the $\chi_{1}$-$\chi_{2}$ plane. The corresponding results are demonstrated in Figs.~\ref{CAR_phase}-\ref{CAR_lem2}. 
The second one is devoted to the discussion of the same via the perfect transmission/CT process. We present the corresponding results
in Figs.~\ref{CT_phase}-\ref{CT_lem2}.

\subsection{Pumping via CAR in the $\chi_{1}$-$\chi_{2}$ plane}
Silicene is a material where a large value of non-local CAR process can be obtained due to its unique band structure~\cite{Linder2014}.  
The band gaps and Fermi level (chemical potential) in silicene can be tuned by applying electric fields only. By tuning the both, very
recently, Linder \etal\, in Ref.~\onlinecite{Linder2014} showed that one can completely block elastic cotunneling 
in silicene NSN junction in the subgapped regime. Consequently, pure CAR process is possible for a broad range of energies. However, maximum 
probability of CAR found in Ref.~\onlinecite{Linder2014} was $\sim 96.2\%$ while the rest was normal reflection probability. 

\begin{figure}[!thpb]
\centering
\includegraphics[width=0.9\linewidth]{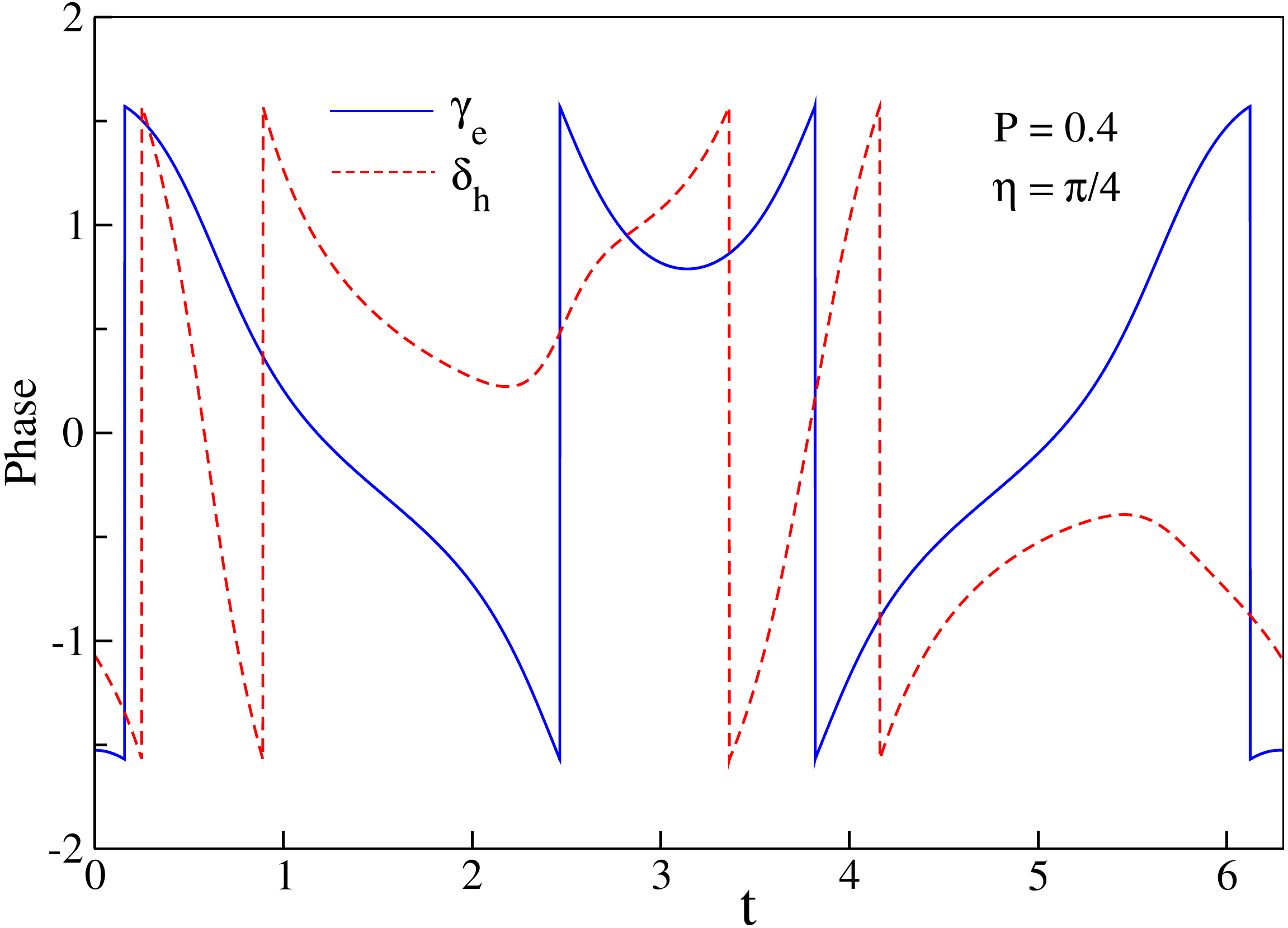}
\caption{(Color online) The plot shows the variation of the normal reflection phase $\gamma_{e}$ and CAR phase $\delta_{h}$, 
with time $t$, along a chosen pumping contour in the $\chi_1$ - $\chi_2$ plane. 
}
\label{CAR_phase}
\end{figure}

\begin{figure}[!thpb]
\centering
\includegraphics[width=1.0\linewidth]{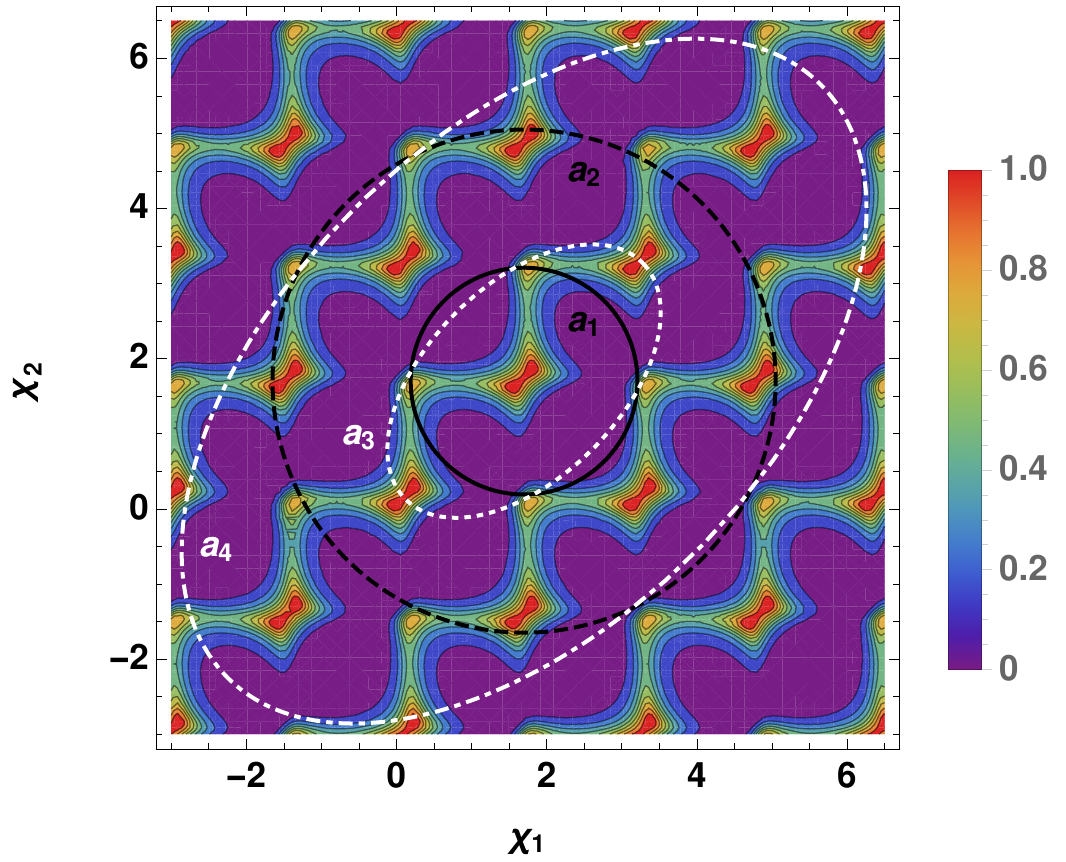}
\caption{(Color online) Plot of CAR probability $|t_{A}|^2$ in $\chi_1$-$\chi_2$ plane. The contours $a_1,\,a_2$ represents $\eta=\pi/4$ 
and $P=1.51$, $P=3.35$ respectively. On the other hand, the contours $a_3,\,a_4$ are for $\eta=\pi/6$ and $P=1.82$, $P=4.56$ respectively. 
The value of the other parameters are chosen to be $L=2.1\xi$, $E=0.9 \Delta$, $\omega=1$, $\chi_0=1.7$, $\mu_L=5 \Delta$, $\mu_R=-5 \Delta$, 
$\mu_S=20 \Delta$ and $\lambda_L=\lambda_R=5 \Delta$.
}
\label{CAR_cir1}
\end{figure}

The probability of non-local CAR process can be enhanced to unity (100\%) (see Fig.~\ref{CAR_cir1}) by introducing two insulating barriers at each 
NS interfaces. We have considered $\mu_L=5 \Delta$, $\mu_R=-5 \Delta$ and $\lambda_L=\lambda_R=5 \Delta$ which reflects the fact that the Fermi 
level touches the bottom of the conduction band in the left normal silicene side while it touches the top of the valance band in right normal 
silicene side. This is illustrated in Fig.~\ref{band}. The superconducting silicene side is doped with $\mu_S=20 \Delta$ to satisfy mean field 
condition for superconductivity $\mu_S\gg\Delta$~\cite{Linder2014}. The band gaps $\lambda_L$ and $\lambda_R$ at the two normal sides can be 
adjusted by the external electric field $E_{zm}$ (m=L/R). The chosen value of the band gaps and doping levels permits one to neglect the contribution 
from the other valley (${\bf{K^{\prime}}}$) which has much higher band gap compared to the other energy scales in the system (see Fig.~\ref{band}). 
Under such circumstances, we obtain pure CAR in this setup choosing length of the superconducting side, $L=2.1\xi$ ($\xi=\hbar v_F/\pi\Delta$
is the phase coherence length of the superconductor) and incident electron energy, $E=0.9 \Delta$. Note that, for our
analysis, we choose the same parameter values as used in Ref.~\onlinecite{Linder2014}.

The reason behind obtaining pure CAR process in our NISIN set-up is as follows. As there is a band gap 
$2\lambda_L=2(elE_{zL}-\tilde{\eta}\sigma\lambda_{SO})>\Delta$ in the left normal 
silicene side, probability for AR to take place is vanishingly small~\cite{Linder2014,surajit2016conductance}. On the other hand, 
$2\lambda_{R}=2(elE_{zR}-\tilde{\eta}\sigma\lambda_{SO})$ is the energy gap between the conduction band and valance band 
in the right (R) normal silicene region as illustrated in Fig. 2. Moreover, the chemical potential $\mu_{R}$ in the right (R) normal silicene 
is chosen to be at the top of the valence band. Hence, only hole states are available in the right normal side. Therefore, an electron incident
from the conduction band of the left normal silicene region encounters a gap and unavailability of electronic states to tunnel into the right 
normal region which essentially block the co-tunneling (CT) probability.
Hence, the only possible scattering processes remain are normal reflection and CAR.
This allows our system to possess completely pure CAR process with 
probability one in $\chi_1\,-\,\chi_2$ plane as shown in Fig.~\ref{CAR_cir1}. These resonant CAR peaks are $\pi/2$ periodic in nature 
and they appear in pairs. Such periodic nature and the fact that resonaces appear in pairs, affect the pumped charge behavior 
which will be discussed later. The oscillatory behavior of the CAR resonance can be explained as follows. Non-relativistic free electrons with 
energy $E$ incident on a potential barrier with height $V_0$ are described by an exponentially decaying (non-oscillatory) wave function inside 
the barrier region if $E < V_0$, since the dispersion relation is $k\sim \sqrt{E-V_0}$. On the contrary, relativistic free electrons satisfies 
a dispersion $k\sim (E-V_0)$, consequently corresponding wave functions do not decay inside the barrier region~\cite{lindersudbo,subhro,
paul2016thermal}. Instead, the transmittance of the junction displays an oscillatory behavior as a function of the strength of the barrier. 
Hence, the undamped oscillatory behavior of CAR is a direct manifestation of the relativistic low-energy Dirac fermions in silicene. 
The periodicity depends on the Fermi wave-length mismatch between the normal and superconducting 
region~\cite{paul2016thermal,surajit2016conductance}.

Note that, the Fermi energy (chemical potential) need neither necessarily exactly touch valance band maxima or conduction band minima nor they 
need to have same magnitude at the two normal regions to obtain resonant CAR. 
A small deviation, from the numerical values that we have taken, also leads to the resonant CAR probability to take place within the subgapped regime. 
Previously, possibility of obtaining CAR was also reported in $p$-$n$ junction of graphene~\cite{cayssol} at a specific value of the parameters. 
However, a small deviation from that leads to CT along with CAR contaminating that possibility.

As phases of the scattering amplitudes play a major role in the determination of the pumped charge, we show the behavior of 
phases of normal reflection and CAR amplitudes ($\gamma_{e}$ and $\delta_{h}$ respectively) as a function of time for one full cycle 
in Fig.~\ref{CAR_phase}. We observe that both $\gamma_{e}$ and $\delta_{h}$ exhibit four abrupt jumps for a full period of time
(along a chosen contour). These jumps play a significant role in determining the pumped charge which we emphasis later. In addition, 
throughout our analysis, we have considered incident electrons to be normal to the interface \ie\, $\alpha_{eL}=0$ for simplicity. 
Later for completeness, we demonstrate angle dependence of the pumped charge.

\begin{figure}[!thpb]
\centering
\includegraphics[width=0.98\linewidth]{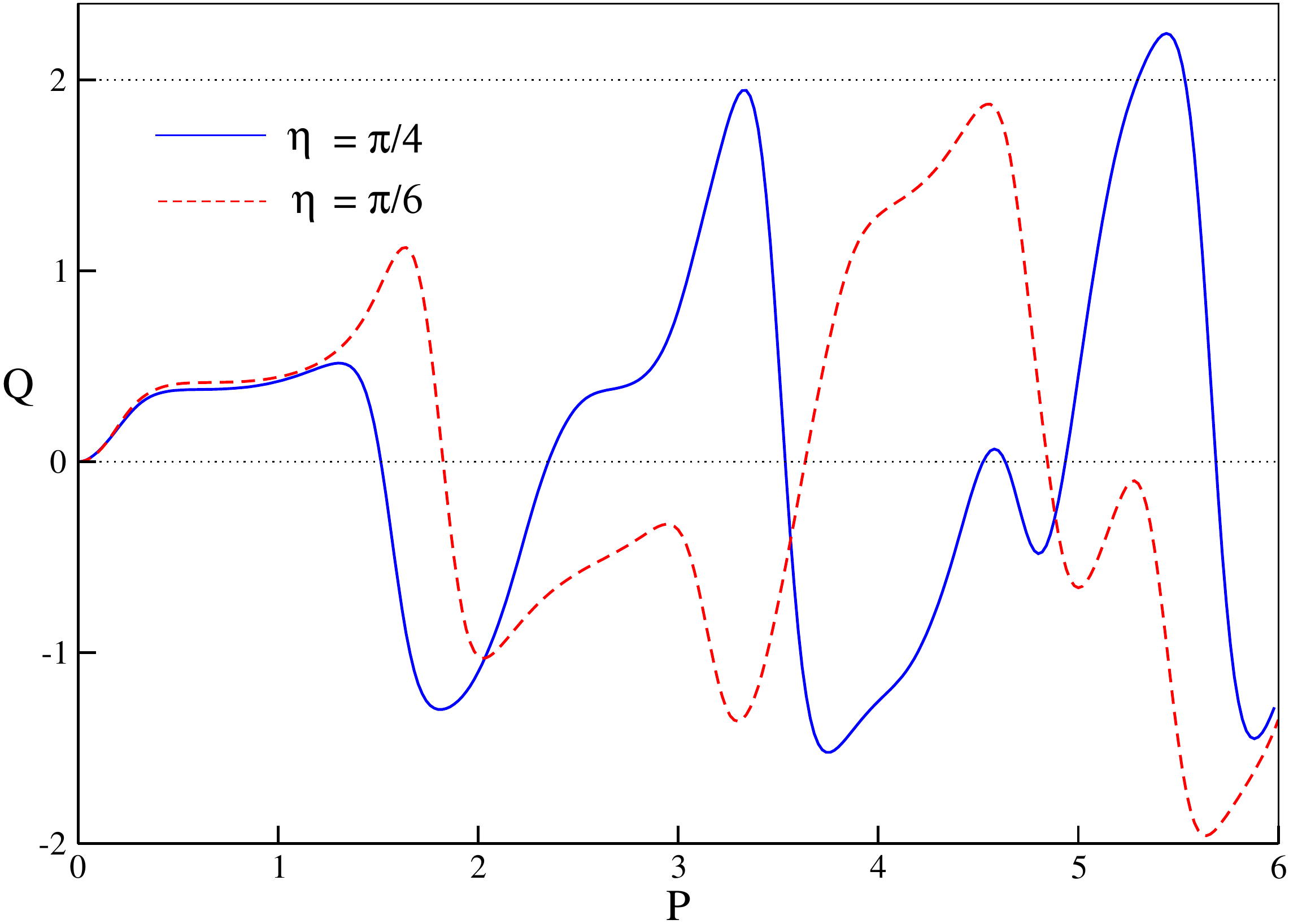}
\caption{(Color online) The pumped charge $Q$ in units of the electron charge $e$, for pumping in the $\chi_1$-$\chi_2$ plane, is shown 
as a function of the pumping strength $P$ for circular and elliptic contours. The value of the other parameters are chosen to be the same
as mentioned in Fig.~\ref{CAR_cir1}.}
\label{CAR_cir2}
\end{figure}

Under such scenario where the only possible scattering processes are normal reflection and CAR, Eq.(\ref{pumpch}) simplifies to 

\begin{eqnarray}
Q=\frac{e}{2\pi}\int_0^{T} dt[&-&|t_A|^2(\dot{\delta}_h \cos{\alpha_{hR}}+\dot{\gamma}_e \cos{\alpha_{eL}})\non \\ 
&+&\dot{\gamma}_e \cos{\alpha_{eL}}]\ ,
\label{EqCAR}
\end{eqnarray}

The behavior of pumped charge $Q$ as a function of the pumping strength $P$ is shown in Fig.~\ref{CAR_cir2} for $\eta=\pi/4,\,\pi/6$ which 
correspond to circular and elliptic contour respectively. The features of $Q$, depicted in Fig.~\ref{CAR_cir2}, can be understood from the 
behavior of CAR probability $|t_{A}|^2$ in the $\chi_1$-$\chi_2$ plane.
For small values of $P$, pumped charge $Q$ becomes vanishingly small in magnitude as the pumping contours do not enclose any $|t_{A}|^2=1$ point. 
When a pumping contour encloses one of the resonant peaks of $|t_{A}|^2$, topological part of the pumped charge gives rise to $ne$ 
($n$ is the winding number) due to the integration around a singular point. At this point the reflection phase $\gamma_{e}$ becomes ill-defined.
However, the dissipative part nullifies the topological part resulting in small values of $Q$ (see Eq.(\ref{pumpch})) for both $\eta=\pi/4,\,\pi/6$.
On the other hand, when a contour encloses both $|t_{A}|^2$ resonances, the relative integration direction around the two singular points 
plays an important role. Namely, when two resonances are enclosed in a path with the same orientation, then the two contributions
have opposite sign and tend to cancel each other.
For \eg~when $\eta=\pi/4$ (black circular contours $a_1$ and $a_2$), the pumped charge is zero for $P=1.51$ (see Fig.~\ref{CAR_cir2})
as the contour $a_1$ encloses both the peaks resulting in zero pumped charge. 
Similar feature was found in case of resonant transmission in Ref.~\onlinecite{lev,entin2002quantized,banerjee2007adiabatic,saha2014quantum} 
where pumped charge was found to be zero when the pumping contour encloses both the resonances. 
$Q$ approaches almost quantized value $2e$ for $P=3.35$ and the corresponding contour $a_2$ encloses even number of resonance pairs in the 
same orientation. Hence the topological part of pumped charge is almost zero and the contribution to $Q$ arises from the dissipative part.
The large contribution from the dissipative part arises due to the total drop of the CAR phase $\delta_{h}$ by a factor of $4\pi$ during its 
time evolution along the contour $a_{2}$ (see Fig.~\ref{CAR_phase}). Similarly, when $\eta=\pi/6$, $Q$ is zero at $P=1.82$ which corresponds 
to the $a_3$ contour which encloses four peaks (two pairs) in total, resulting in zero contribution from the topological part. On the other hand, 
pumped charge reaches its maximum when $P=4.56$ ($a_4$ contour) where also the entire contribution originates from the dissipative part 
(see Fig.~\ref{CAR_cir2}). Pumped charge $Q$ exceeds the value $+2e$ as pumping strength $P$ increases (see Fig.~\ref{CAR_cir2}) 
for both $\eta=\pi/4$ and $\pi/6$. Physically, the contribution of the dissipative part in pumped charge increases non-monotonically with 
the pumping strength. Hence, as the pumping contour encloses more number of pairs of resonant CAR peaks, due to the enhancement of dissipative part, 
pumped charge can exceed +2e with further increase of $P$.
Pumped charge can change sign depending on the sense of enclosing of the resonances \ie~whether it is clock-wise or anti-clockwise. 
 
\begin{figure}[!thpb]
\centering
\includegraphics[width=0.99\linewidth]{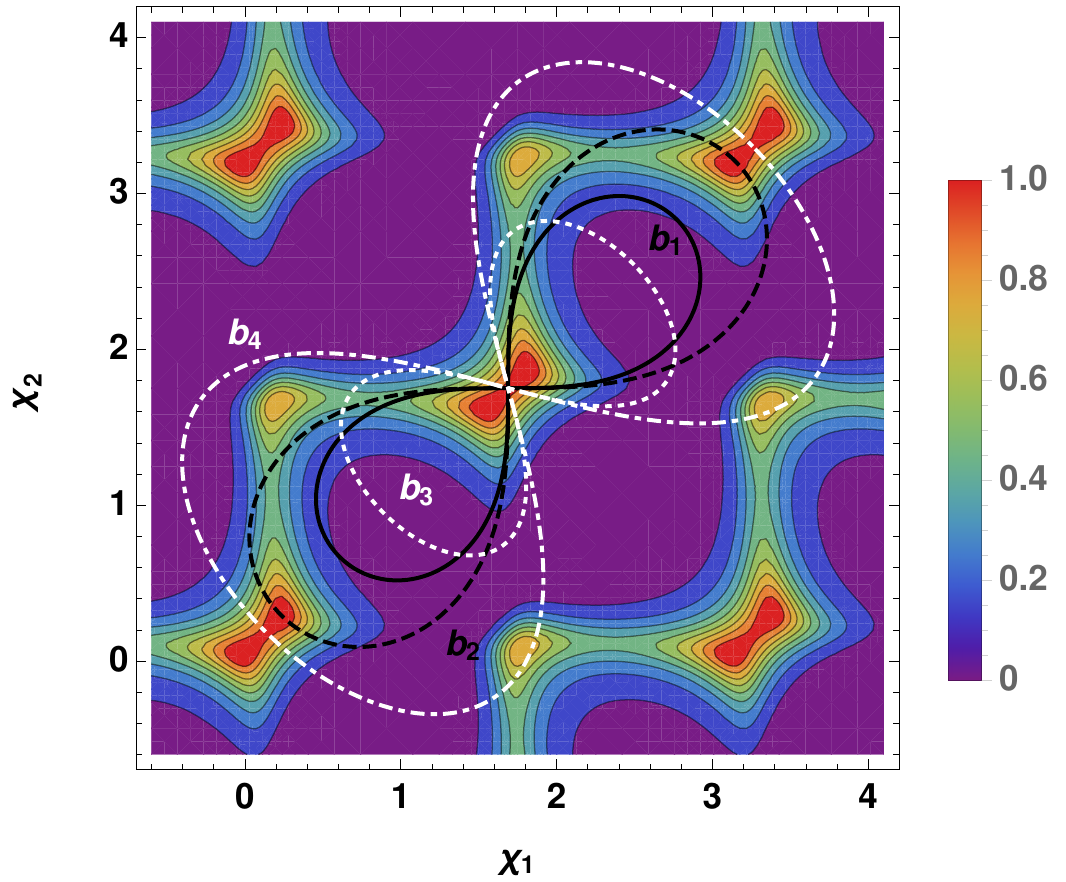}
\caption{(Color online) Plot of CAR probability $|t_{A}|^2$ along with lemniscate contours are shown in the $\chi_1$-$\chi_2$ plane. 
The contours $b_1,\,b_2$ represents $\theta=\pi/4$ and the contours corresponding to $\theta=\pi/3$ are $b_3,\,b_4$. 
We have chosen the mean values $\chi_{1_0}=1.69$ and $\chi_{2_0}=1.75$. The value of the other parameters are chosen to be the same
as mentioned in Fig.~\ref{CAR_cir1}.}
\label{CAR_lem1}
\end{figure}

\begin{figure}[!thpb]
\centering
\includegraphics[width=0.98\linewidth]{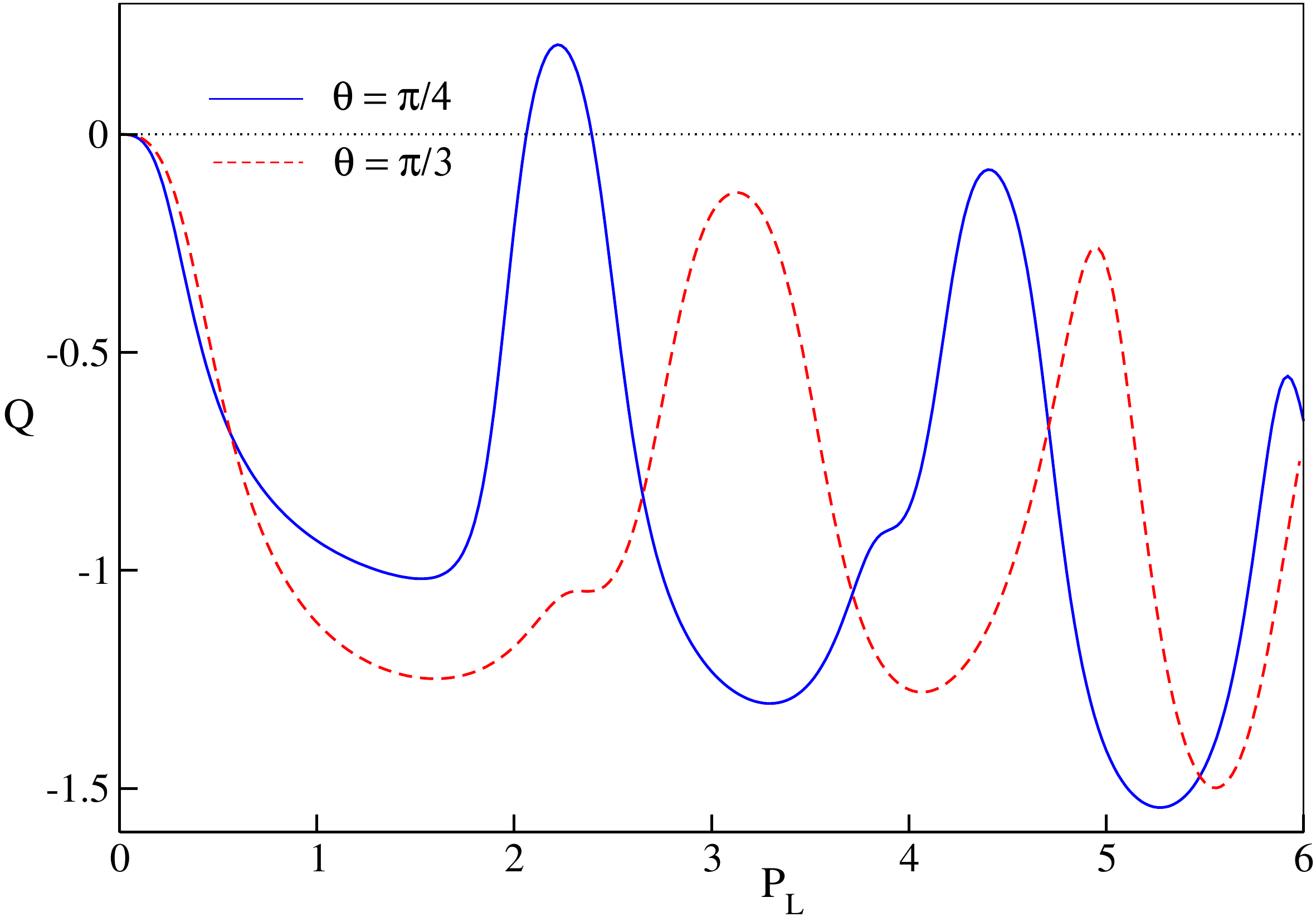}
\caption{(Color online) Pumped charge $Q$ in unit of electon charge $e$, for pumping in the $\chi_1$-$\chi_2$ plane, is shown as a function of 
the pumping strength $P_L$ for the lemniscate contours. All other parameters are identical to those used in Fig.~\ref{CAR_cir1}.}
\label{CAR_lem2}
\end{figure}

The behavior of pumped charge $Q$ with respect to the pumping strength $P_{L}$ for lemniscate contours with $\theta=\pi/4$ and $\pi/3$ is 
presented in Fig.~\ref{CAR_lem2} and the corresponding contours are shown in Fig.~\ref{CAR_lem1}. The pumped charge is small for small 
values of $P_L$ where the contribution from topological part is cancelled by the dissipative part. As $P_{L}$ increases, the corresponding 
pumping contour encloses both the $|t_{A}|^2$ peaks within opposite integration orientations and as a consequence, the two contributions for
the pumped charge sum up. This is exactly the reason that motivates us to choose the lemniscate contours.
 However, the dissipative part effectively reduces the total pumped charge. Such feature arises for lemniscate contours of the type $b_1$ and $b_3$.
Moreover, we observe that the pumped charge becomes zero for $P_L=2.06$ at $\theta=\pi/4$, where both the bubbles of the $b_2$ contour enclose 
two $|t_{A}|^2$ peaks from the two adjacent resonances in the $\chi_1$-$\chi_2$ plane and hence their combined contribution to pumped charge 
get cancelled for each bubble separately. The qualitative behavior of $Q$ remains similar for $\theta=\pi/3$ where maximum value of $Q$ is achieved 
when each bubble of the lemniscate contour of type $b_4$ encloses odd number of resonance pairs while $Q$ tends to zero as even number of pairs are 
enclosed by each bubble of the contour. 


\subsection{Pumping via transmission/CT in the $\chi_{1}$-$\chi_{2}$ plane}
In this subsection we present our numerical results for the adiabatic quantum pumping through pure CT \ie~resonant transmission process. 
The latter can be achieved by tuning the Fermi level (chemical potential) at the bottom of the conduction band in both the normal silicene regions
(see Fig.~\ref{band}). The numerical values of all the parameters are identical to those used before except now $\mu_R=5\Delta$, $L=2.2\,\xi$ 
and $E=0.93\Delta$. 

\begin{figure}[!thpb]
\centering
\includegraphics[width=0.92\linewidth]{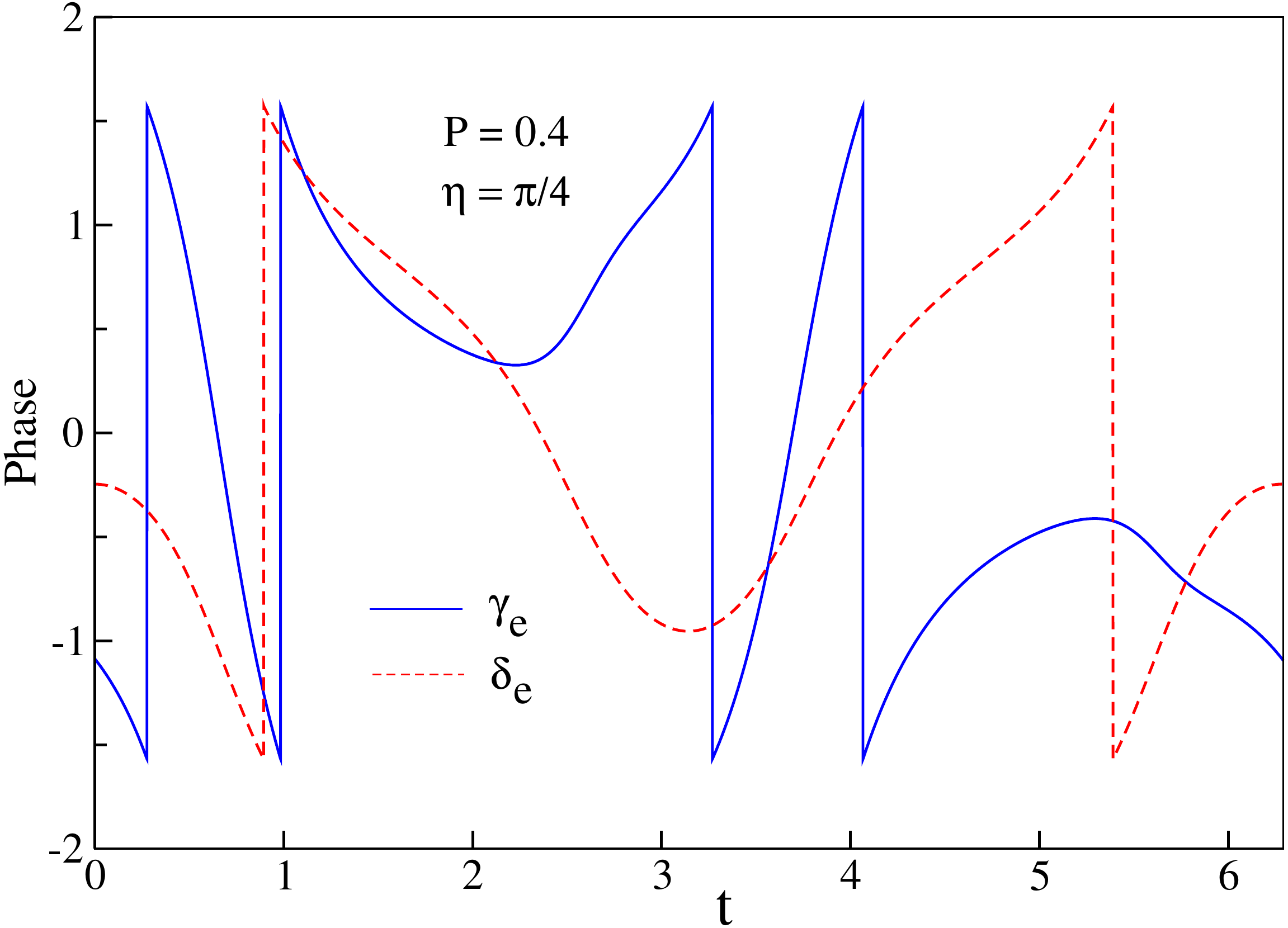}
\caption{(Color online) The variation of the normal reflection phase $\gamma_{e}$ and transmission phase $\delta_{e}$, 
with time $t$, is shown along a chosen pumping contour in the $\chi_1$ - $\chi_2$ plane.
}
\label{CT_phase}
\end{figure}

\begin{figure}[!thpb]
\centering
\includegraphics[width=1.0\linewidth]{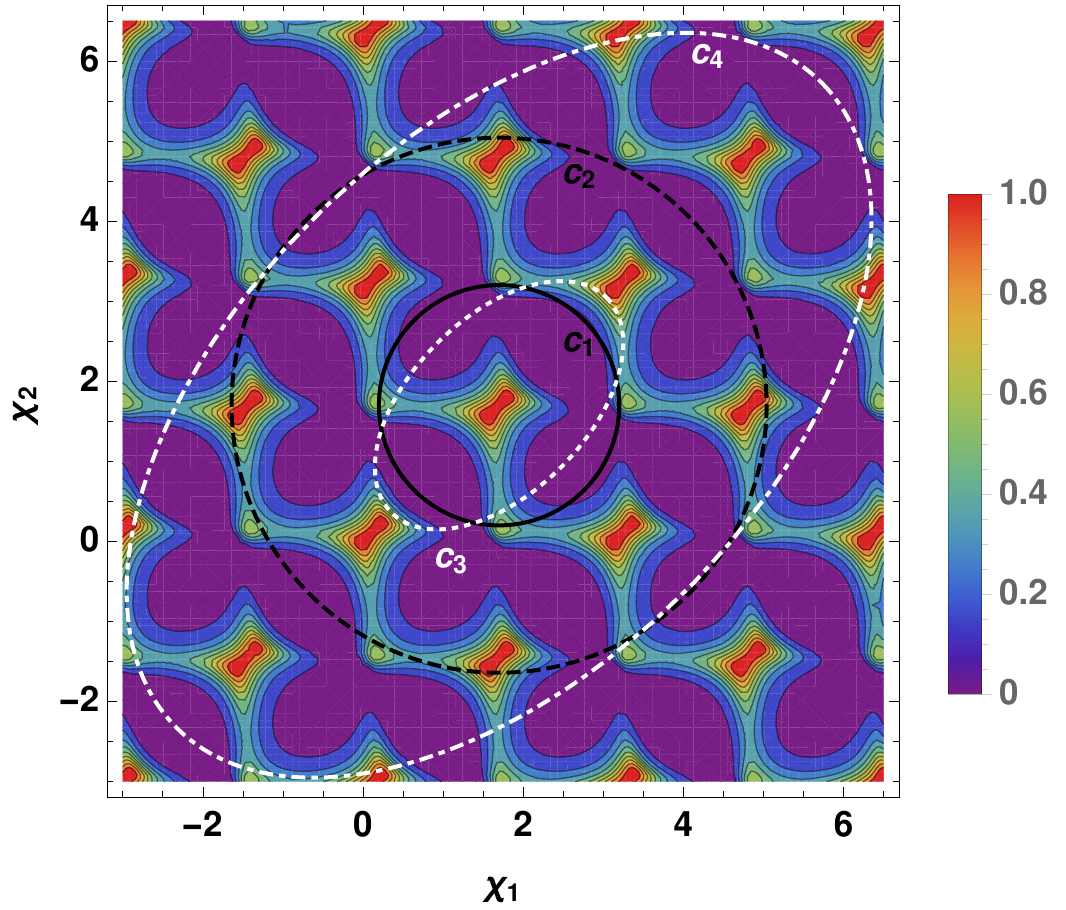}
\caption{(Color online) Transmission probability $|t_{e}|^2$ along with circular and elliptic contours are shown in $\chi_1$-$\chi_2$ plane. 
The contours $c_1,\,c_2$ represent $\eta=\pi/4$ and $P=1.5$, $P=3.34$ respectively. On the other hand, the contours $c_3,\,c_4$ 
correspond to $\eta=\pi/6$ and $P=1.55$, $P=4.65$ respectively. The value of the other parameters are chosen to be $L=2.2\xi$, 
$E=0.93\Delta$, $\omega=1$, $\chi_0=1.7$, $\mu_L=5 \Delta$, $\mu_R=5 \Delta$, $\mu_S=20 \Delta$ and $\lambda_L=\lambda_R=5 \Delta$.
}
\label{CT_cir1}
\end{figure}

As before, due to the presence of a gap (2$\lambda_{L}>\Delta$) in the left normal side, AR is forbidden while CAR cannot take place because 
of the unavailability of the hole states in the right normal region in the low energy limit. An incident electron thus only encounters two 
scattering processes which are normal reflection and transmission. The presence of insulating barriers between the NS interfaces allows both 
these scattering probabilities to be oscillatory as a function of the dimensionless barrier strengths $\chi_1$ and $\chi_2$ which is depicted 
in Fig.~\ref{CT_cir1}.

In this regime, as AR and CAR probabilities are always zero, hence Eq.(\ref{pumpch}) reduces to

\begin{eqnarray}
Q&=&\frac{e}{2\pi}\int_0^{T} dt[|t_{e}|^2(\dot{\delta}_e \cos{\alpha_{eR}}-\dot{\gamma}_e \cos{\alpha_{eL}})\non\\ 
&+&\dot{\gamma}_e \cos{\alpha_{eL}}]\ ,
\label{EqCT}
\end{eqnarray}

\begin{figure}[!thpb]
\centering
\includegraphics[width=0.97\linewidth]{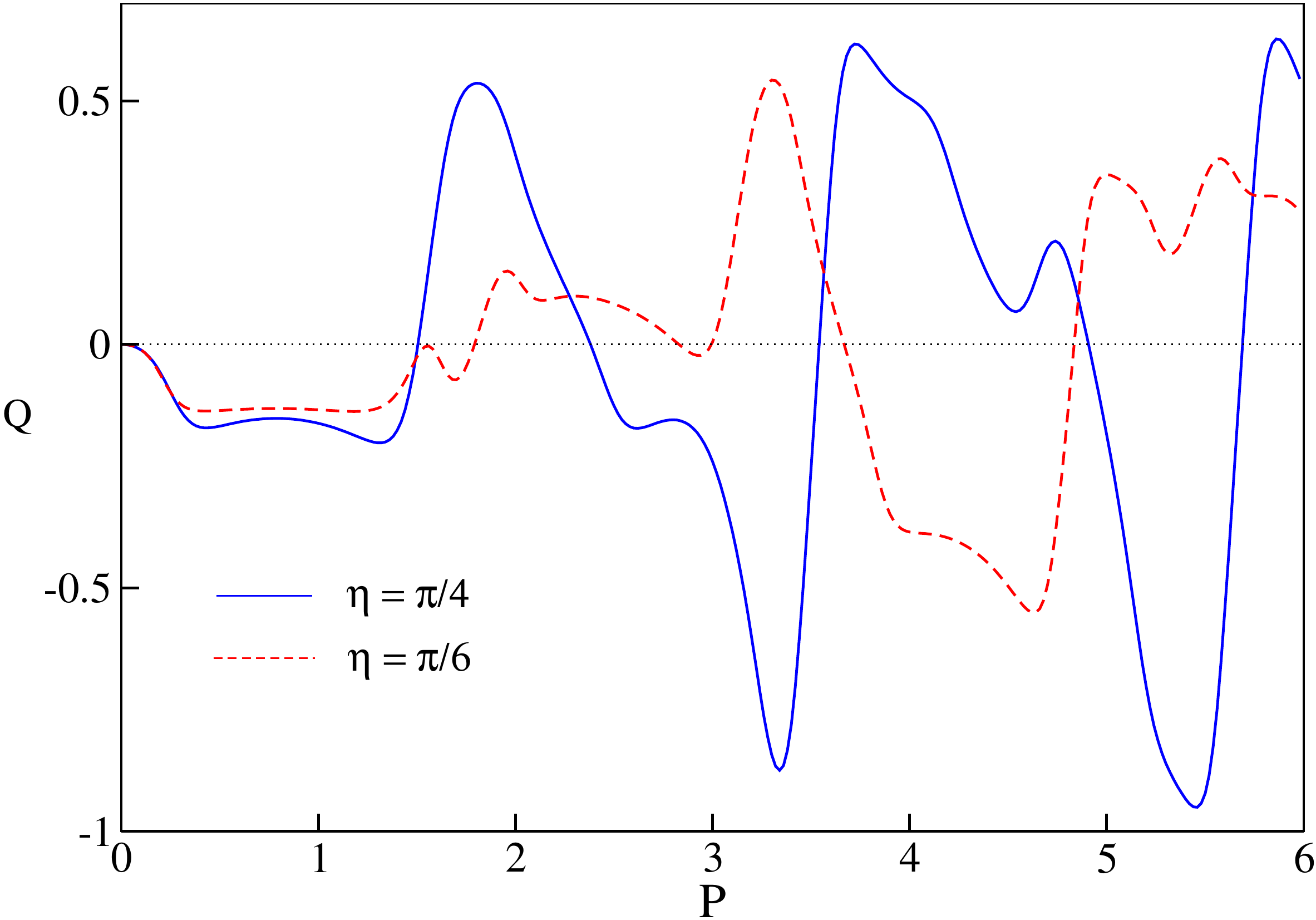}
\caption{(Color online) Pumped charge $Q$ in unit of electron charge $e$, for pumping in the $\chi_1$-$\chi_2$ plane, is shown as a function 
of the pumping strength $P$ for the circular and elliptic contours. We choose the same values of the other parameters as mentioned 
in Fig.~\ref{CT_cir1}.}
\label{CT_cir2}
\end{figure}

In Fig.~\ref{CT_cir2}, pumped charge $Q$ is presented as a function of pumping strength $P$ for $\eta=\pi/4$ (circular contour)  
and $\pi/6$ (elliptic contour). To understand the behavior of the pumped charge, we also investigate the transmission probability 
$|t_{e}|^2$ in $\chi_1-\chi_2$ plane (see Fig.~\ref{CT_cir1}). 
We observe qualitatively similar features of the pumped charge as depicted in the previous subsection.
Here also topological part of pumped charge becomes zero when pumping contour encloses even number of resonance pairs in the same orientation. 
Finite contribution from dissipative part, in $Q$, emerges due to the total jump of the transmission phase $\delta_{e}$ by a factor of $2\pi$ during 
its time evolution along the contour $c_{2}$ (see Fig.~\ref{CT_phase}). On the other hand, for contour $c_{1}$, dissipative part vanishes because 
over a full period of time, reflection and transmission phases $\gamma_{e}$ and $\delta_{e}$ respectively cancell each other (see Eq.(\ref{EqCT})). 
Although, $Q$ approaches to $-e$ for pumping via resonant CT process compared to $2e$ via the resonant CAR process.

\begin{figure}[!thpb]
\centering
\includegraphics[width=1.0\linewidth]{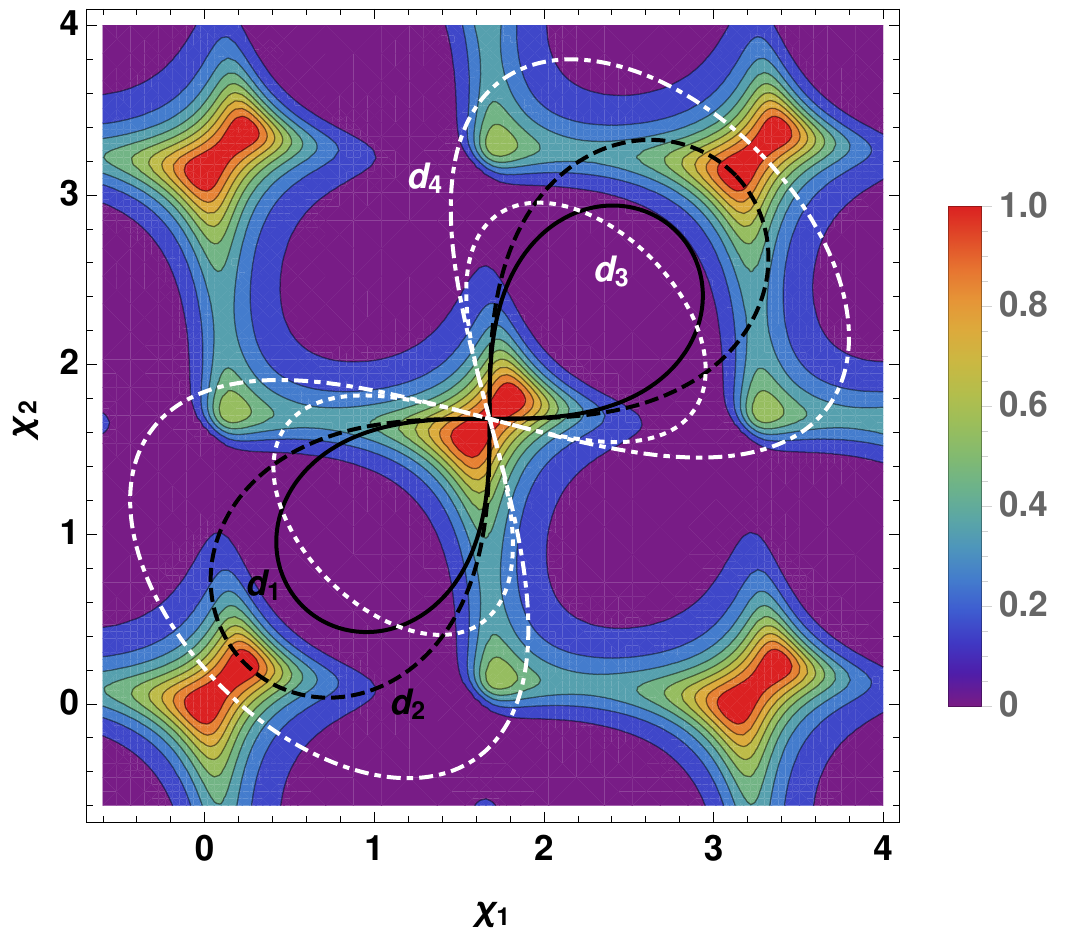}
\caption{(Color online) Transmission probability $|t_{e}|^2$ together with different lemniscate contours are shown in the $\chi_1$-$\chi_2$ plane. 
The contours $d_1,\,d_2$ represents $\theta=\pi/4$ and the contours $d_3,\,d_4$ corresponds to $\theta=\pi/3$. We choose the values of 
$\chi_{1_0}$ and $\chi_{2_0}$ as $\chi_{1_0}=\chi_{2_0}=1.68$. All other parameters are identical to those used in Fig.~\ref{CT_cir1}.
}
\label{CT_lem1}
\end{figure}

\begin{figure}[!thpb]
\centering
\includegraphics[width=0.97\linewidth]{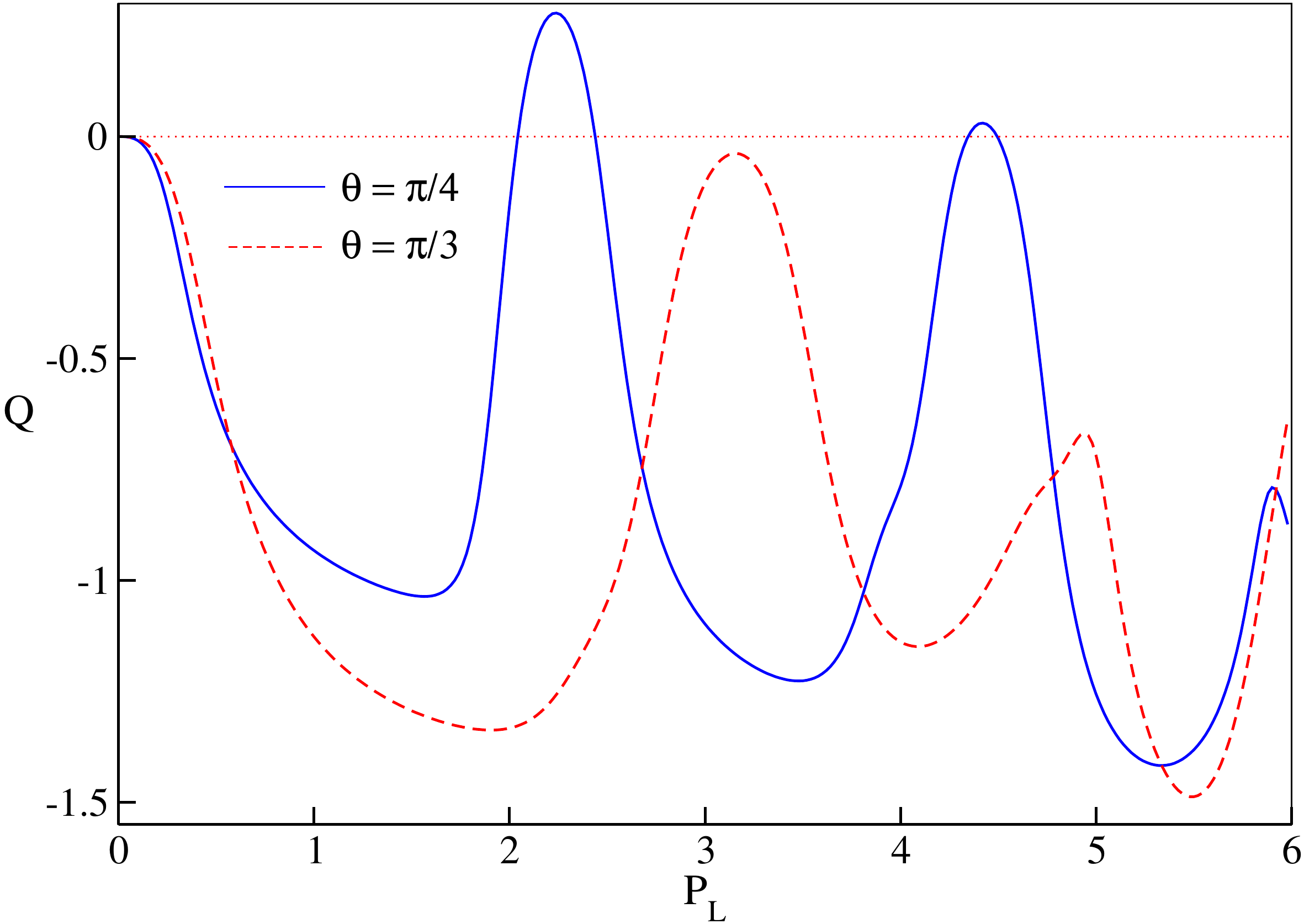}
\caption{(Color online) Pumped charge $Q$, in unit of electron charge $e$, is depicted as a function of the pumping strength $P_L$ for the 
lemniscate contours. All other parameters are identical to those used in Fig.~\ref{CT_cir1}.}
\label{CT_lem2}
\end{figure}

In Fig.~\ref{CT_lem2}, we show the behavior of pumped charge $Q$ as a function of the pumping strength $P_{L}$ with lemniscate contours. 
To understand the corresponding behavior of $Q$, we also show $|t_{e}|^2$ in the $\chi_{1}$-$\chi_{2}$ plane along with different lemniscate 
contours (see Fig.~\ref{CT_lem1}). Here also the features of $Q$ remains similar as previous subsection for both $\theta=\pi/4$ and $\pi/3$.

\begin{figure}[!thpb]
\centering
\includegraphics[width=0.9\linewidth]{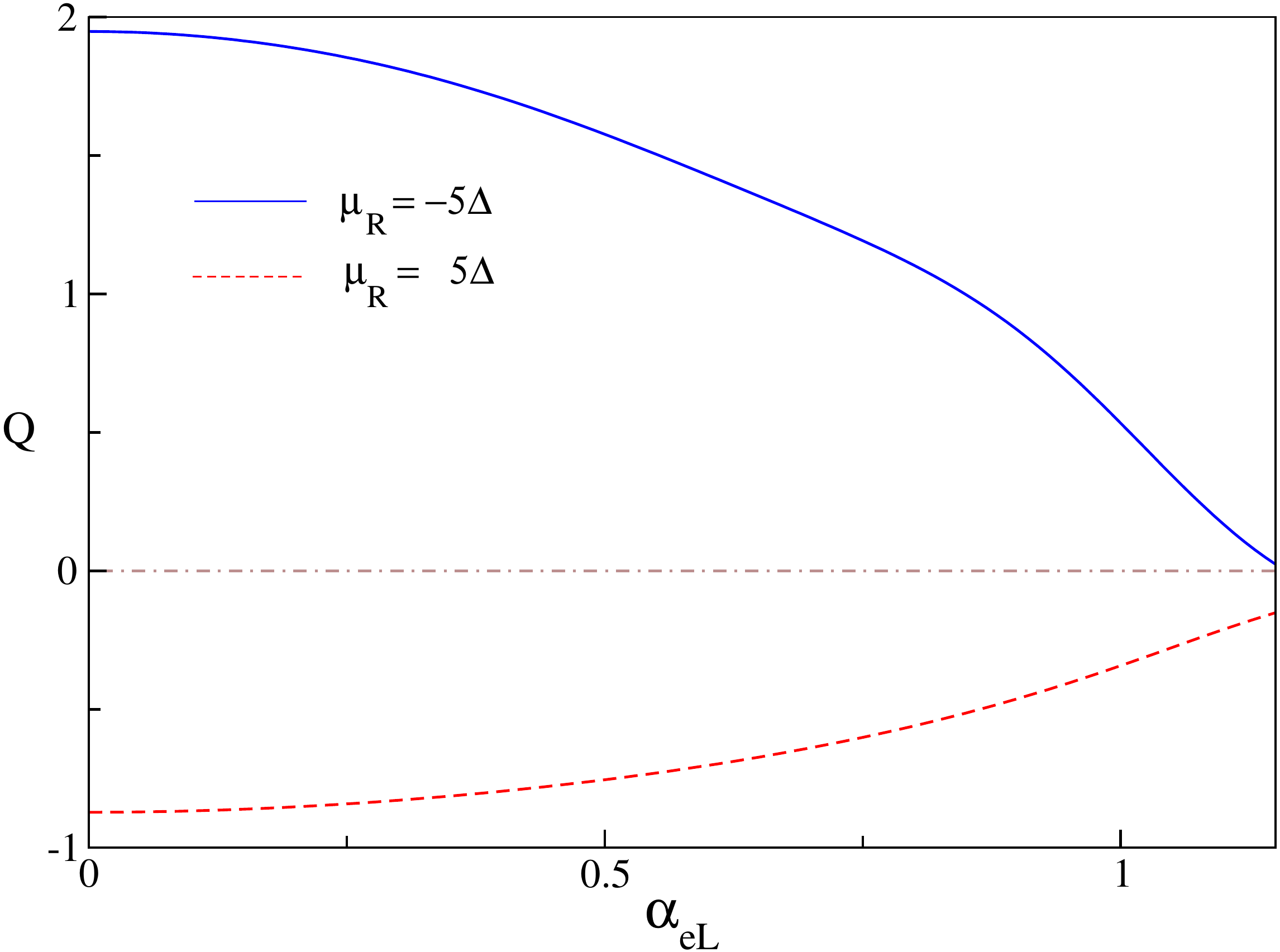}
\caption{(Color online) Pumped charge $Q$, in unit of electron charge $e$, is shown as a function of the incident angle $\alpha_{eL}$ 
for both $\mu_R=-5\Delta$ and $\mu_R=5\Delta$. Here we choose $\eta=\pi/4$, $P=3.35$ for $\mu_R=-5\Delta$ and $P=3.34$ for 
$\mu_R=5\Delta$ respectively. 
}
\label{angle}
\end{figure}

As we mention earlier, the above mentioned results are valid for normal incidence of the incoming electron \ie\, $\alpha_{eL}=0$. 
Here, we explore the dependence of the pumped charge on the angle of incident electrons. In Fig.~\ref{angle}, pumped charge $Q$ as a function 
of incident angle $\alpha_{eL}$ is presented when either CAR probability $|t_{A}|^2$ or transmission probability $|t_{e}|^2$ is enclosed by the 
circular pumping contour. 
The $\alpha_{eL}$ dependence is shown upto the critical angle $\alpha_c$. Above $\alpha_c$, AR and CAR processes cannot take place~\cite{beenakker1}. 
Rather, normal reflection is the dominating scattering mechanism above $\alpha_c$. It is evident from Fig.~\ref{angle} that as the angle of 
incidence $\alpha_{eL}$ increases, $Q$ decreases monotonically for enclosing $|t_{A}|^2$ or $|t_{e}|^2$ in either cases. The reason can be attributed 
to the fact that both $|t_{A}|^2$ and $|t_{e}|^2$ in the two different scenarios, acquire the maximum value at normal incidence \ie\, $\alpha_{eL}=0$ 
and decreases slowly with the increase of $\alpha_{eL}$. Also, for $0<\alpha_{eL}<\alpha_c$, normal reflection probability $|r_{e}|^2$ also 
contributes to Eq.(\ref{pumpch}) and the interplay between all the quantum mechanical amplitudes and their phases results in smaller value of 
pumped charge.
Note that, in case of pumping via CAR resonance process in $\chi_1-\chi_2$ plane, $Q$ approaches zero as $\alpha_{eL}$ proceeds towards $\alpha_c$. 
However, $Q$ is finite even at $\alpha_c$ in case of pumping via resonant transmission in the same parameter space, 
This is because at $\alpha_c$, $|t_{A}|^2$ vanishes while $|t_{e}|^2$ still has small probability which gives rise to small pumped
charge arising from the dissipative part (see Eq.(\ref{EqCT})). 

\section{Summary and conclusions} {\label{sec:IV}}
To summarize, in this article, we have investigated the possibility of enhancing the CAR probability $|t_{A}|^2$ in silicene NSN set up by 
introducing thin insulating barrier~\cite{paul2016thermal,surajit2016conductance} $I$ at each NS interface. We show that, for electrons with 
normal incidence, resonant CAR can be obtained in our setup by tuning the band gap in both the normal silicene regions by applying 
an external electric field as well as adjusting the chemical potential by additional gate voltages. We also show 
that $|t_{A}|^2$ is periodic in $\chi_1$-$\chi_2$ plane due to relativistic nature of Dirac fermions.
On the other hand, it is also possible to attain transmission probability $|t_{e}|^2$ of magnitude unity in silicene NISIN junction under
suitable circumstances. Owing to Dirac nature of particles, $|t_{e}|^2$ also exhibits periodic behavior in the space of barrier 
strengths $\chi_1$ and $\chi_2$. 

We then explore adiabatic quantum charge pumping through our NISIN setup and show that the behavior of pumped charge as a function of the pumping
strength $P$ is closely related to the features of CAR probability $|t_{A}|^2$ or transmission probability $|t_{e}|^2$ in the pumping parameter space. 
For electrons with normal incidence, large pumped charge with value close to $Q\sim2e$ can be obtained when 
particular circular or elliptic pumping contour encloses the resonant CAR in $\chi_1$-$\chi_2$ plane. Although the major contribution to $Q$, in this 
case, arises from the dissipative part. On the other hand, large pumped charge can also be obtained with lemniscate contour when odd number of 
$|t_{A}|^2$ peaks are enclosed by each of its bubble. In contrast, pumped charge approaches to $Q\sim-e$ when various pumping contours enclose 
$|t_{e}|^2$ resonance in the same parameter space. However, pumped charge decreases monotonically as we increase the angle of incidence of the 
incoming electron. In experimental situation, the measurable quantity should be the angle averaged pumped charge analogous to angle averaged
conductance~\cite{andy2016exp}. From our analysis, we expect that the qualitative nature of angle averaged pumped charge as a function of the 
pumping strength will remain similar to the $\alpha_{eL}=0$ case. Although the quantitative value of $Q$ will be smaller than the angle resolved 
case as $Q$ decreases monotonically with $\alpha_{eL}$. 

Note that, our calculation is valid for zero temperature. Nevertheless, in our case, temperature 
$T_{p}$ must be smaller than the proximity induced superconducting gap $\Delta$. We expect that the qualitative features of our results 
for the pumped charge will survive in the presence of low temperatures. For non-zero yet small temperatures, $T_{p} \ll \Delta$, the 
pairs of resonant peaks in the parameters space will have a slight broadening due to thermal smearing. Therefore, we believe that the 
qualitative features of pumped charge $Q$ with respect to the pumping strength $P$ will still be captured in our model. Although there 
can be quantitative change in $Q$. On the other hand, if $T_{p}>\Delta$, then CAR process from the interface will decay and pumped charge
will become vanishingly small due to thermal fluctuation.

As far as practical realization of our silicene NISIN quantum pumping set up is concerned, superconductivity in silicene may be possible to induce 
by proximity coupled to a $s$-wave superconductor for \eg~$\rm Al$, $\rm NbSe_{2}$ analogous to graphene~\cite{heersche2007bipolar,
choi2013complete,andy2016exp}. Once such proximity induced superconductivity in silicene 
is realized, fabrication of silicene NISIN junction can be feasible. The strength of the two oscillating barriers can be 
possible to tune by applying a.c gate voltages. Typical spin-orbit energy in silicene is $\lambda_{\rm SO}\sim 4~\rm meV$
and the buckling parameter is $l\approx 0.23~\rm \AA$~\cite{liu2011low,ezawa2015}. Considering Ref.~\onlinecite{heersche2007bipolar,
calado2015ballistic}, typical proximity induced superconducting gap in silicene would be $\Delta \sim 0.2~\rm meV$. For such induced superconducting 
gap, chemical potential is $\mu_S\,\sim\,20\Delta\,\sim\,4~\rm meV$ and we obtain $\xi\,\sim\,580~\rm nm$ and length of the superconducting region 
$L\sim 1.2~\rm \mu m$. Hence, an insulating barrier of thickness $d\sim 10-20~\rm nm$ may be considered as thin barrier and the gate voltage 
$V_{0}\sim 500~\rm meV$ can therefore justify the needs of our model~\cite{paul2016thermal}. To achieve both the resonances, 
$\lambda_{L}=\lambda_{R}=5\Delta\sim 1~\rm meV$ which can be tuned by an external electric field $E_{zL}=E_{zR}\sim 200~\rm V/\mu m$. For both 
resonant processes, typical dwell time of the electrons inside the superconducting region is $\sim 2.2\, \rm fs$ while the time period of 
the oscillating barriers is $T\sim 30\,\rm ps$ and the corresponding frequency of modulation parameters turns out to be $\sim 230\, \rm GHz$. 
Thus the dwell time $\tau_{dwell}$ is much smaller than the time period $T$ of the modulation parameters, hence satisfying the adiabatic condition 
of quantum pump. Pumped current through our setup should be in the range of $\sim10-15\, \rm nA$ which can be measurable in 
modern day experiment.

\bibliography{Pumping_ref} 

\end{document}